\documentclass[aps,prd,twocolumn,showpacs,showkeys,nofootinbib,groupedaddress]{revtex4-1}

\usepackage{graphicx}
\usepackage{dcolumn}
\usepackage{bm}
\usepackage{amsmath}
\usepackage{amsfonts,amssymb}
\usepackage{color}
\usepackage{enumerate}
\usepackage{float}
\usepackage{subfig}
\usepackage{ulem}
\usepackage{times}

\begin{document}

\title{Dynamical cosmologies in Eddington-inspired-Born-Infeld theory}

\author{N. M. Jim\'enez Cruz$^{1}$}
\email{nmjc1209@gmail.com}

\author{Celia Escamilla-Rivera$^{2}$}
\email{celia.escamilla@nucleares.unam.mx}

\affiliation{$^1$ Facultad de Ciencias en F\'isica y Matem\'aticas, Universidad Aut\'onoma de Chiapas. \\ 
Ciudad Universitaria, Carretera Emiliano Zapata Km. 8, Real del Bosque (Ter\'an), 29050, Tuxtla Guti\'errez, Chiapas, M\'exico.}

\affiliation{$^2$ Instituto de Ciencias Nucleares, \\ Universidad Nacional Aut\'onoma de M\'exico, 
Circuito Exterior C.U., A.P. 70-543, M\'exico D.F. 04510, M\'exico.}

%%%%%%%%%%%%%%%%%%%%%%%%%%%%%%%%%%%%%%%%%%%%%%%%%%%%
%%%%%%%%%%%%%%%%%%%%%%%%%%%%%%%%%%%%%%%%%%%%%%%%%%%%

\begin{abstract}
In this paper we study the cosmological evolution of the universe filled with a perfect fluid in the Eddington-inspired Born-Infeld gravity. Applying an alternative method in which the evolution of the scale factor for this theory is linked to the cosmographic parameters, we obtain a dynamical dark energy solution where the singularity (through a regular bounce or a loitering phase) still can be avoided for $\kappa>0$ with $w>0$. For the range $-1<w<0$, the results lead us to universes that experience an unlimited rate of expansion with finite density. Also, we obtain a possible maximum value of $H=4.338\times10^{-43}/t_{\text{Planck}}$ at the cosmic bounce point.
\end{abstract}

%\keywords{Cosmology; Modified theories of gravity.}
%\pacs{98.80.+k; 04.50.Kd}
\maketitle

%%%%%%%%%%%%%%%%%%%%%%%%%%%%%%%%%%%%%%%%%%%%%%%%%%%%
%%%%%%%%%%%%%%%%%%%%%%%%%%%%%%%%%%%%%%%%%%%%%%%%%%%%

\section{Introduction}
\label{sec:intro}
Einstein's Theory of General Relativity (GR) with Cosmological Constant ($\Lambda$) is, so far, the best theory that describes the behaviour of space-time at macroscopic scales, unfortunately fails at scales where energies are close to $10^{19}$GeV. 
In this limit is where a new theory of gravity is necessary, due that one still seeks to describe in a simple way the evolution of the universe and its components. 
However, the possibility and demand for revised GR is also coming to the fore. This arises from its own problem of GR or is required from newly found observations. Currently, there are two big problems in that regards: \textit{(1)} The theory includes the existence of singularity, which denies the application of GR itself, as a solution for field equations. \textit{(2)} The rising of the problems of dark energy and dark matter. 

In the light of rich observed data, it is possible to uncover some properties in order to found a solution for the above problems. In any case, we can consider each component of the dark sector or one might have a new proposal of the gravitational theory without the need of these dark components instead. Some attempts has been done in order to achieve these issues, e.g in \cite{Banados:2008fi} was presented a class of bigravity with solutions that can be interpolate between matter and acceleration epochs. In this model it is possible to describe a coherent unified dark sector where an Eddington Born-Infeld action (EBI) represent a connection
between the fundamental fields, so-called the affine connections, and the rest of the universe through the gravitation.  A remarkable point of this theory is the presence of two metrics, one of them $g_{\mu\nu}$ is the usual metric that couples with matter. The second metric, $q_{\mu\nu}$,
satisfies the Einstein-Hilbert action and couples to the rest of the  world through its interaction with $g_{\mu\nu}$. This theory is view as a bimetric model in where $q_{\mu\nu}$ can be interpreted as dark matter and not as an extra scalar field. The resulting cosmology show that the EBI scenario can mimic the standard cosmological evolution, where the field can behave as pressureless matter and $\Lambda$ at the same time. These theories of bigravity are extensively 
study in the literature \cite{Damour:2002ws,ArkaniHamed:2002sp,Blas:2005yk,Banados:2008fj} and reference therein.
All these models precise an affine connection equivalent to Einstein gravity. To avoid this preference in \cite{Banados:2010ix} was presented a non-conventional formulation in terms of the affine connection $\Gamma^{\mu}_{\alpha\beta}$ (which is assumed to be symmetrised) and
a space-time metric $g_{\alpha\beta}$ such that the gravitational action is given by:

\begin{equation}
S_{\text{EiBI}}=\frac{2}{\kappa}\int d^{4}x\left[\sqrt{|g_{\mu \nu}+\kappa R_{\mu \nu}(\Gamma)|}-\lambda\sqrt{-g}\right]+S_{M}. \label{EiBI}
\end{equation}
where $\lambda$ is a non-zero scalar, dimensionless. $g$ is the determinant of the metric $g_{\mu\nu}$, $|g_{\mu\nu}+\kappa R_{\mu\nu} (\Gamma)| $ is the determinant of $g_{\mu\nu}+\kappa R_{\mu\nu} (\Gamma)$, $R_{\mu \nu} (\Gamma) $ represents the symmetric part of $R_{\mu \nu}$ that is built with the $\Gamma$ connection and $S_M$ is the usual matter action. The case in which $\lambda =0$ is meaningless when doing the variation of $S_{\text{EiBI}}$ with respect to the metric.

For small values of $\kappa R$ in Eq.\eqref{EiBI}, one gets the action of Einstein-Hilbert with $\Lambda=(\lambda-1)/\kappa$. Whereas, for large values of $\kappa R$, the action approaches that of Eddington.
Under the Palatini formalism by vary $S_{EiBI}$ respect to the metric and the connection, respectively, one obtains
\begin{eqnarray}
\sqrt{\frac{|q|}{|g|}}q^{\mu \nu}-\lambda g^{\mu \nu}&=&-\kappa T^{\mu \nu}, \label{eq:1} \\
\nabla_{\alpha}(\sqrt{q} q^{\mu \nu})&=&0.\label{eq:2}
\end{eqnarray}
It was convenient to introduce an auxiliary metric $q_{\mu \nu}=g_{\mu \nu}+ \kappa R_{\mu \nu}$ compatible with the connection. Since its proposal, EiBI theory has many cosmological implications, e.g. it was found that some scenarios containing baryonic matter avoid singularities depending the sign of $\kappa$ and by considering universes which experience unbounded expansion rate \cite{Scargill:2012kg}. Some cosmological studies of this theory has been carried out in order to study structure formation at large scales \cite{Du:2014jka}, using a perfect fluid \cite{Cho:2012vg}, or at local regimes where the gravitational properties of dark matter halos are present \cite{Harko:2013xma}. Also, linear perturbations has been done for scenarios in where $k$-modes can gave an unstable behaviour near the minimum scale in both regular bounce and at the loitering phase \cite{EscamillaRivera:2012vz,Yang:2013hsa,Cho:2014jta}.
 
In this paper we not only obtain the expected behaviours above described by using another method to resolve the field equations, as an extension we investigate cosmographic quantities in order to describe the dynamics behind the theory. For this purpose, the structure is presented as follow: in Sec. \ref{sec:cosmoEiBI} we cover the basic field equations of the EiBI theory and describe each cosmological implication. As a new result, we found an oscillatory behaviour of the scale factor involved in the theory. In Sec. \ref{sec:deceleration_p} we present the exact solution for the deceleration parameter either in the asymptotic past or at the bounce with a cosmographic approach. Finally, in Sec. \ref{sec:conclusions} we discuss our final results.

%%%%%%%%%%%%%%%%%%%%%%%%%%%%%%%%%%%%%%%%%%%%%%%%%%%%
%%%%%%%%%%%%%%%%%%%%%%%%%%%%%%%%%%%%%%%%%%%%%%%%%%%%

\section{EiBI background cosmology}
\label{sec:cosmoEiBI}

\begin{figure}[ht!]
 \centering
    \includegraphics[width=0.45\textwidth]{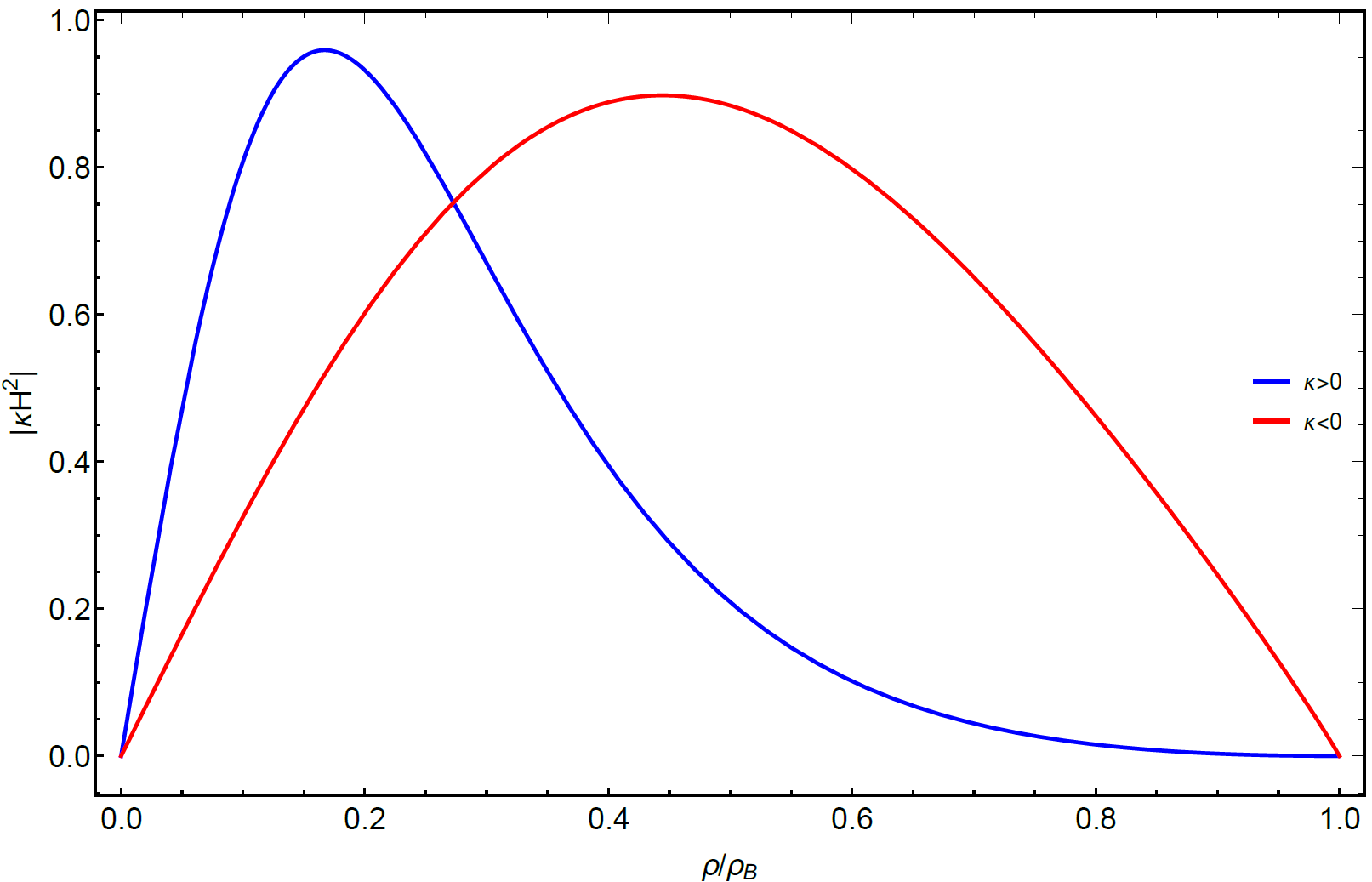}
 \caption{Evolution of the Hubble parameter in EiBI theory. Blue-line (red-line) corresponds to the case $\kappa>0$ ($\kappa<0$), respectively. The density was normalised by  $\rho_{B}$ ($\rho_{B}=3/\kappa$ for $\kappa>0$ and $\rho_{B}=-1/\kappa$ for $\kappa<0$). 
}
 \label{f:Hubbleb}
\end{figure}

Field equations (\ref{eq:1})-(3) can be solved analytically using the traditional homogeneous and isotropic metric as 
a line element with time and spatial components:
\begin{eqnarray}\label{eq:metrics1}
g_{\mu\nu} dx^{\mu} dx^{\nu}&=& -a^2 d\eta^2 +a^2 dx^2,\\
q_{\mu\nu} dx^{\mu} dx^{\nu} &=& -X^2 d\eta^2 +Y^2 dx^2,
\end{eqnarray}
in that way we can derive the conventional Friedmann cosmology at late-times. 
The zero-component evolution equation with $\sqrt{|q/g|}= |XY^{3}|$ is: 
\begin{equation}\label{eq:friedmann}
 3\kappa \left(H +\frac{\dot{Y}}{Y}\right)^2 =X^2 \left(1-\frac{3}{2Y^2}\right)+\frac{1}{2},
\end{equation}
where
\begin{eqnarray}\label{eq:X-Y}
 |X|&=&\frac{(1+\kappa P_{T})^{2}}{[(1+\kappa \rho_{T})(1-\kappa P_{T})^{1/4}]},\\
  |Y|&=&[(1+\kappa \rho_{T})(1-\kappa P_{T})^{1/4}],
\end{eqnarray}
with $\rho_{T}=\rho +\Lambda$ and $P_{T}=P-\Lambda$. 
Let us assume radiation domination as: $\rho_{T}=\rho$
and $P_{T}=P=\rho/3$, we find that $X$ and $Y$ at late times behaves as:
\begin{eqnarray}
 |X|&\simeq& 1-\frac{5}{6}\kappa\rho+O(\kappa^2), \\
  |Y|&\simeq& a+\frac{a}{6}\kappa\rho+O(\kappa^2).
\end{eqnarray}
If $X=Y=1$ the latter reduces to the low-energy densities limit (GR limit). Now, considering high energy densities
(Eddington limit) $\rho\rightarrow \rho_{B}$, where the subindex $B$ 
indicates the existence of a minimum value for the scale factor \cite{Banados:2010ix} then 
the approximation for the variables $X$ and $Y$ are
\begin{eqnarray}
 |X|&=&\frac{\left(1-\frac{\bar{\rho}}{3}\right)^2}{[(1+\bar{\rho})\left(1-\frac{\bar{\rho}}{3}\right)]^{1/4}}, \\
  |Y|&=&[(1+\bar{\rho})\left(1-\frac{\bar{\rho}}{3}\right)]^{1/4},
\end{eqnarray}
where we introduce $\bar{\rho}=\kappa\rho$. Notice that two critical point at $\bar{\rho}=\rho_{B}=3$ for $\kappa >0$
and $\bar{\rho}=\rho_{B}=-1$ for $\kappa <0$.
Rewriting Eq.(\ref{eq:friedmann}) we obtain
\begin{eqnarray}\label{eq:friedmann1}
 3H^2 &=&\frac{1}{\kappa}\left[\bar{\rho} -1 +\frac{1}{3\sqrt{3}}\sqrt{(\bar{\rho} +1)(3-\bar{\rho})^3}\right] 
 \nonumber \\ && 
\times \left[\frac{(1+\bar{\rho})(3-\bar{\rho})^2}{(3+\bar{\rho}^2)^2}\right],
\end{eqnarray}
where for $\bar{\rho} \ll 1$ we have $H^2 \simeq \rho/3$. Eq.(\ref{eq:friedmann1}) has critical points 
for $H(\rho_{B})=0$ in a maximum density $\rho_{B} =0,-1,3$. Each critical point appear when 
$Y^2 =3X^2 /(2X^2 +1)$. Each critical density has an analytical solution 
that corresponds to an expansion of the scale factor depending of the sign of $\kappa$ (see Figure \ref{f:Hubbleb}):
\begin{itemize}
 \item When $\bar{\rho} \rightarrow 0$, 
 $X=Y=1$, then we have a minimum scale factor at $\dot{a}=0$ and the universe its stationary and has a
minimum size $a=a_{B}\approx 10^{-32}(\kappa)^{1/4}a_{0}$, where $a_{0}$ is the scale factor today.
 \item When $\bar{\rho}=3 (\kappa >0)$, $X=Y=0$ and the
solution is exponential-like $(a/a_{B})-1 \propto e^{t-t_{B}}$, which corresponds to a loitering solution.
 \item When $\bar{\rho}=1$ ($\kappa <0$), 
 $X=(3\cdot 4^3)^{1/4}/9$ and
$Y=(4^3)^{1/4}$, with solution $(a-a_{B})\propto |t-t_{B}|^2$, which corresponds to a bouncing solution.
This replace the usual Big Bang singularity of Einstein's model by a cosmic bounce.
\end{itemize}

Given that the solution for the radiation is $\rho= \rho_0/a^4=\rho_0 /(a_B +\delta a)^4$, we can expand the density
around the small variation of $a=a_B+\delta$ with $\delta a << a_B$ as
\begin{eqnarray}
\rho &\propto& \rho_B + 4\rho_B \frac{\delta a}{a_B} + O(\delta a^2),
\end{eqnarray}
where $\rho_B =\rho_0 a^{-4}_B$ is the maximum density. As $a=a_B +\delta a$ then $(a/a_B)-1 =\delta a/a_B$,
\begin{eqnarray}\label{eq:density-scale}
\rho &\propto& \rho_B + 4\rho_B \left(\frac{a}{a_B}-1\right) + O\left[a^{2}_B  \left(\frac{a}{a_B}-1\right)^2\right], \nonumber \\
a &\propto& 1+(t-t_B) + O[(t-t_B)^2]. 
\end{eqnarray}
At early times Eq.(\ref{eq:density-scale}) shows a universe with a maximum density and constant scale factor.

So far, we have considered that $w=P/\rho$ is still valid in this theory, for the case of radiation we have
\begin{equation}
\rho(t)=\frac{\rho_{0}}{a^{4}},
\end{equation} 
where $\rho_{0}$ is the value of the radiation density in the present. Therefore, a maximum finite density $\mathbf{\rho_{B}}$ sets a value for the minimum scale factor \textbf{$a_{B}$} non zero given by $a_{B}=(\rho_{0}/\rho_{B})^{1/4}$. This result then indicates a scenario where the universe has a minimum size before develop an expansion phase like in GR.

Now, if we rewrite Eq. \eqref{EiBI} with the metric tensors $g _{\mu\nu}$ and $ q_{\mu \nu}$ under the variational metric formalism and considering the time-time component we have the following dynamical equation:
\begin{equation}
D\left(-\frac{a}{a+3\ddot{a}\kappa}\right)+\lambda=\kappa \rho,
\end{equation}
where $D=\sqrt{|q| / |g|}=XY^{3}$. Then we obtain a highly nonlinear equation:
\begin{equation}\label{eq00}
-Da+\lambda(a+3\ddot{a}\kappa)=\kappa \rho (a+3\ddot{a}\kappa),
\end{equation}
or
\begin{equation}\label{eq4}
3(\lambda - \kappa \rho)\kappa\ddot{a}+(\lambda-\kappa \rho -D)a=0.
\end{equation}
As a simple case, let us consider $D=\textrm{constant}$, the resulting equation is a simple harmonic oscillator with frequency $\omega=\sqrt{B/A} $, where $A=3(\lambda -\kappa \rho) \kappa$, and $B =\lambda- \kappa \rho -D$. Its solution is given by
\begin{equation}\label{eq5}
a(t)=\alpha \cos{\omega t}+\beta\sin(\omega t),
\end{equation}
with $\alpha$ and $\beta$ as constants to be determined given a specific set of initial conditions.
Thus, the Friedmann equation can be written for this case as
\begin{equation}
\frac{\dot{a}}{a}=\frac{\omega(\beta \cos{\omega t}-\alpha\sin{\omega t})}{\alpha \cos{\omega t}+\beta\sin(\omega t)}.
\end{equation}
In GR, $X=Y=1$, $D=1$, $\Lambda=0$ and $\lambda=1$, the frequency of this oscillator is
\begin{equation}
\omega=\left[\frac{\lambda-\kappa \rho -D}{3(\lambda - \kappa \rho)\kappa}\right]^{1/2}=\left[\frac{ \rho}{3(\kappa \rho-1)}\right]^{1/2}.
\end{equation}
If we take into account the following initial conditions:
\begin{eqnarray}
&&a(t_{0})=\alpha \cos{\omega t_{0}}+\beta\sin(\omega t_{0})=1,\\
&&\dot{a}(0)=\omega \beta \cos{(0)}=\omega \beta=0,
\end{eqnarray}
we get 
\begin{equation}
\beta=0 \quad \text{and} \quad \alpha=\frac{1}{\cos(\omega t_{0})}.
\end{equation}
Thus
\begin{equation}
a(t)=\frac{\cos{\omega t}}{\cos{\omega t_{0}}}, \quad \text{and} \quad H=-\omega\tan{\omega t}. \label{eq7}
\end{equation}

\begin{figure}[ht!]
 \centering
    \includegraphics[width=0.45 \textwidth]{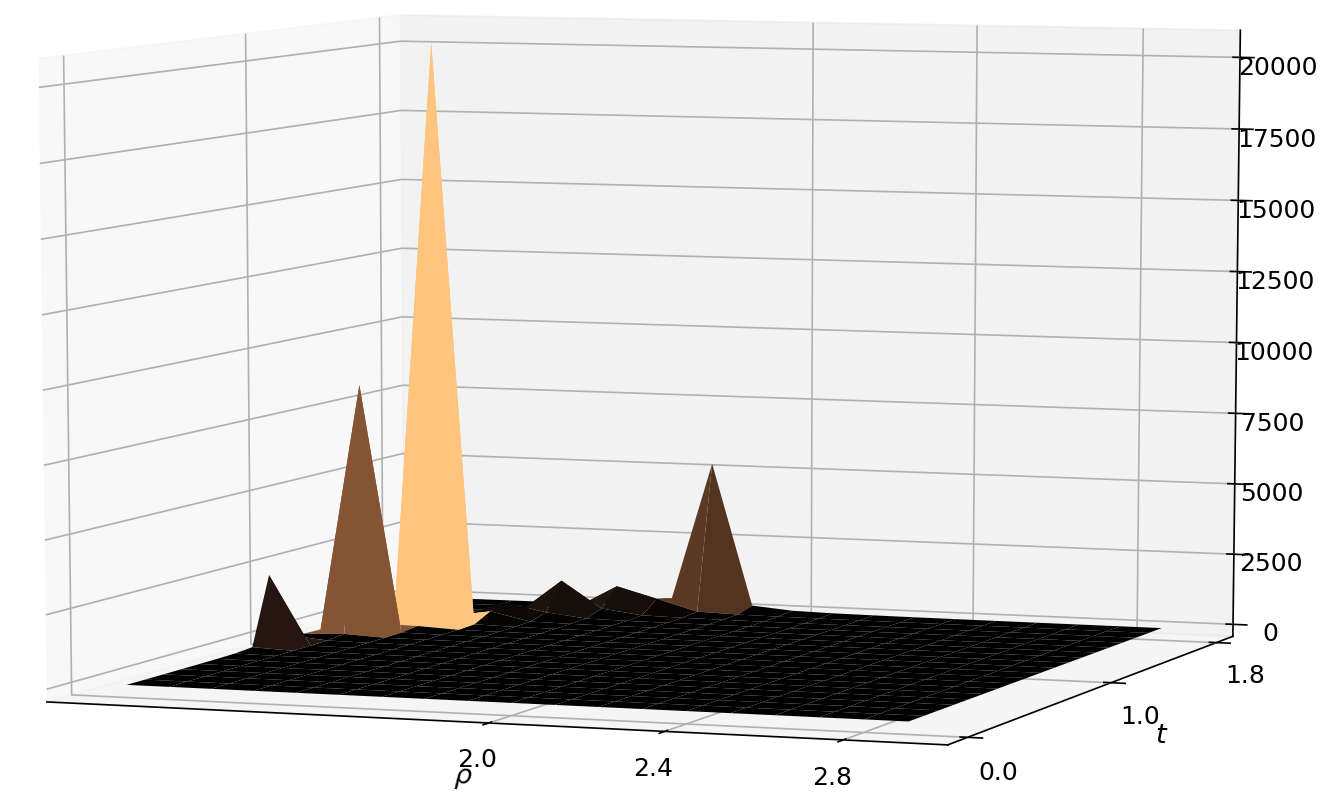}
 \caption{Evolution of $H(t,\rho)^{2}$ between the critical points $\bar{\rho}=1$ and $\bar{\rho}=3$.  Notice that for densities values at the range $\bar{\rho}=1,2$ there are five posible solutions for the scale factor.}
 \label{f:Soluciones1}
\end{figure}

\begin{figure}[ht!]
 \centering
    \includegraphics[width=0.45 \textwidth]{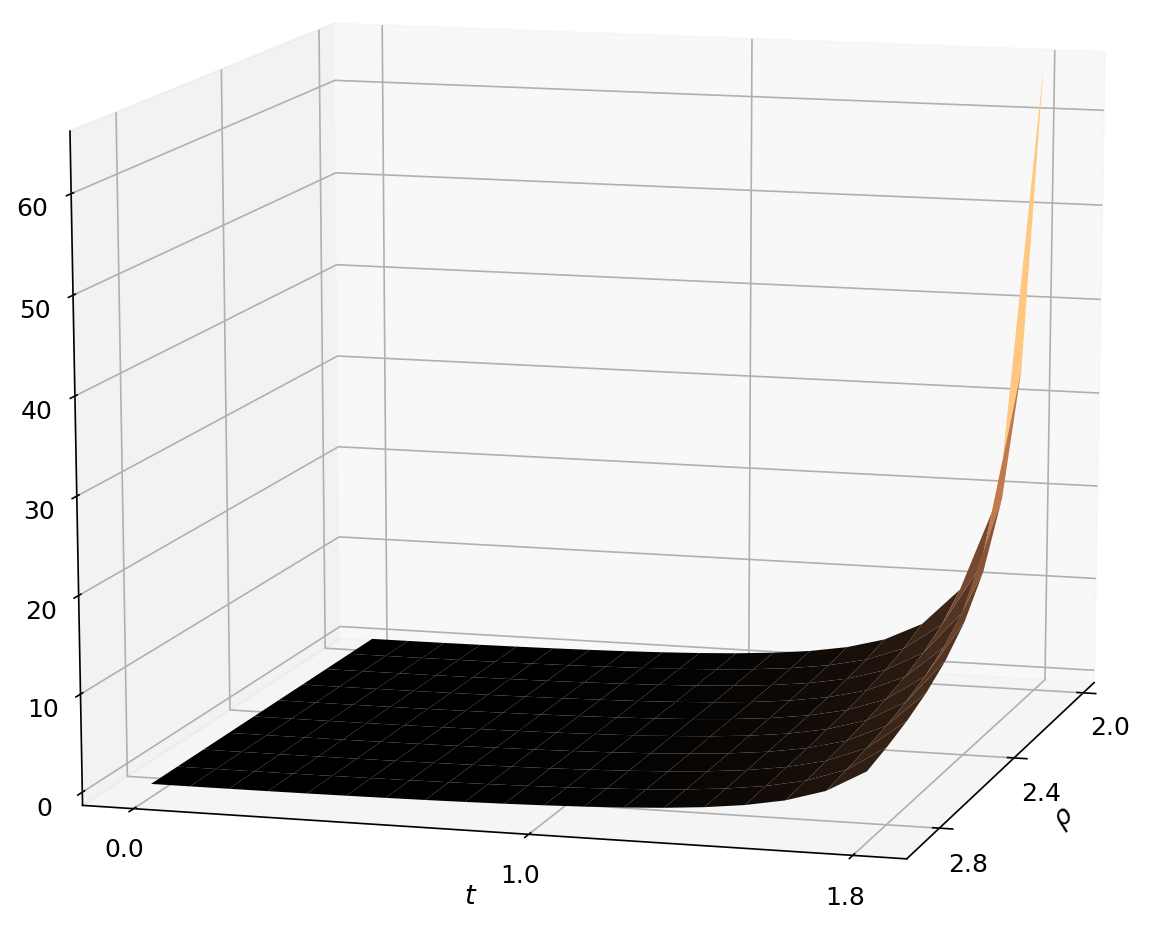}
 \caption{Evolution of $H(t,\rho)^{2}$ between points $\bar{\rho}=2$ and $\bar{\rho}=3$. We notice the maximum value for the scale factor at $\bar{\rho}=2.5$.}
 \label{f:Soluciones2}
\end{figure}

These results show behaviours in which at early times the universe is dominated by radiation, $H=0$, indicating that $\dot{a}=0$ and then $a= \textrm{const}$. However, instead of considering $q_{\mu \nu}$ as an auxiliary metric, this procedure is rather an approximation to the metric $g_{\mu \nu}$ and the result is obtained under a standard variational formalism. The evolution of Eq. \eqref{eq7} for $\kappa=[1,-1]$ is showed in Figures \ref{f:Soluciones1}-\ref{f:Soluciones2}.

We notice how each of the singularities, the loitering phase and the cosmic bounce, still preserves at the density critical points. As an extension of this results, in \cite{Bojowald:2001xe} was showed that a cosmological singularity can be removed by quantum geometry. Alternative, before the cosmic bounce the universe may been in an almost unimaginable quantum state, not yet spacelike, when something triggered the bounce point and the formation of the atoms of spacetime. In such case, a proposal made in \cite{Ashtekar:2009mm} set that using loop quantum gravity (LQC) the big bang is replaced by a quantum bounce which is followed by a robust phase of super inflation where an estimate $H_{\text{Ashtekar}}=0.94/t_{\text{Planck}}$ for the maximum reached by the Hubble parameter is obtained. Comparing our results with the latter, our maximum value at $\bar{\rho}=2.5$ is $H_{\text{EiBI}}=4.338\times10^{-43}/t_{\text{Planck}}$ which correspond to $10^{42}$ order of magnitude less in comparison to $H_{\text{Ashtekar}}$ at the cosmic bounce point. 

%%%%%%%%%%%%%%%%%%%%%%%%%%%%%%%%%%%%%%%%%%%%%%%%%%%%
%%%%%%%%%%%%%%%%%%%%%%%%%%%%%%%%%%%%%%%%%%%%%%%%%%%%
\section{A simple example for EiBI cosmography }
\label{sec:deceleration_p}

With the results obtained from the Hubble parameter through the Palatini formalism, we can obtain the deceleration equation using the acceleration equation by deriving $H$ with respect to time
\begin{equation}
\dot{H}=\frac{3\kappa(\rho +p)}{2Y}\left[\frac{Q}{G}-\frac{3P}{2Y^{3}F}\right]H^{2},
\end{equation}
where
\begin{eqnarray}
P & = & \frac{(\kappa \rho_{T}+\kappa p_{T})(1-\omega-\kappa \rho_{T}-\kappa p_{T})(\kappa p_{T})-1}{Y^4} 
\nonumber \\ && 
+\frac{\omega(1+\kappa\rho_T)}{Y^4} +(1+\omega)(1-\omega-2\kappa\rho_{T}-2\kappa p_{T}), \quad\quad  \\ 
Q & = & \frac{X^{2}}{Y^{3}}\left(\frac{3}{Y^{2}}\right)\left[\kappa p_{T}-1+\omega(1+\rho_{T})\right] \nonumber \\&&
+\omega\left(4X-\frac{6X}{Y^{2}}\right).
\end{eqnarray}
thus
\begin{equation}
\frac{\ddot{a}}{a}=\left[\frac{3\kappa(\rho +p)}{2Y}\left(\frac{Q}{G}-\frac{3P}{2Y^{3}F}\right)+1\right]H^{2},
\end{equation}
Therefore, we can calculate the deceleration parameter as 
\begin{equation}\label{desB}
q=-\frac{\ddot{a}}{a}\frac{1}{H^{2}}=-\left[\frac{3\kappa(\rho +p)}{2Y}\left(\frac{Q}{P}-\frac{3Q}{2Y^{3}F}\right)+1\right],
\end{equation}

Again, if we consider only radiation, the deceleration parameter have the following form

\begin{eqnarray} \label{eq:q}
q &=& -\frac{2 \bar{\rho} \left[\sqrt{-3 \bar{\rho}^{2}+6 \bar{\rho}+9} \bar{\rho}+3 \left(\sqrt{-3 \bar{\rho}^{2}+6 \bar{\rho}+9}-6\right)\right]}{(\bar{\rho}+1) \left[9 \bar{\rho}+\sqrt{3} \sqrt{-(\bar{\rho}-3)^3 (\bar{\rho}+1)}-9\right]}
\nonumber \\&&   
+\frac{8 \bar{\rho} [\bar{\rho} (\bar{\rho}+4)-1]}{(\bar{\rho}-3) (\bar{\rho}+1) \left(\bar{\rho}^{2}+1\right)^2} -1
\end{eqnarray}

\begin{figure}[ht!]
 \centering
    \includegraphics[width=0.4\textwidth]{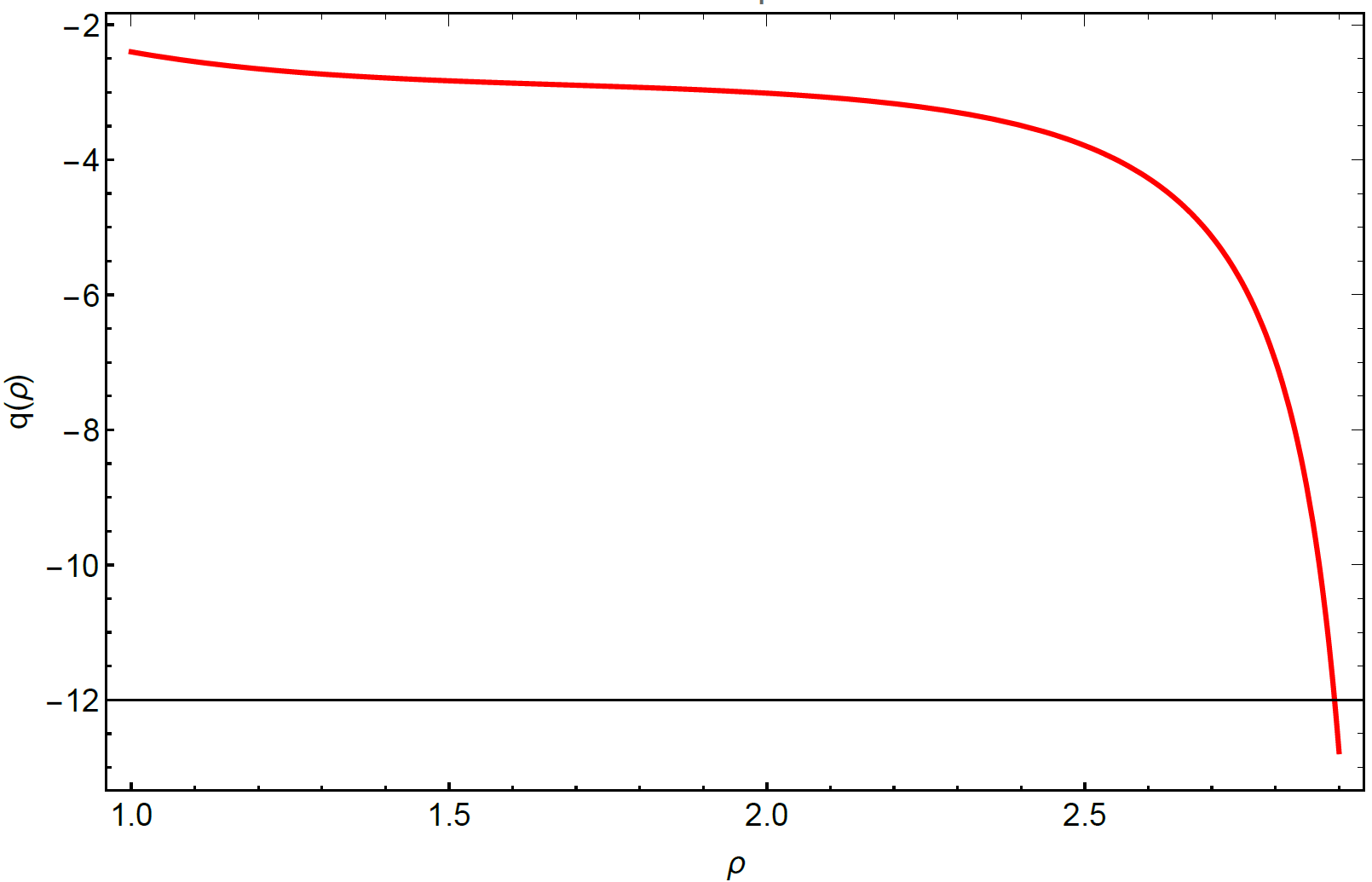}
 \caption{Evolution of Eq. (\ref{eq:q}) for $\kappa=1$ at the critical points $\bar{\rho}=1$ and $\bar{\rho}=3$. Observe that the value of $q$ is always negative, indicating that the cosmic expansion is accelerating in the loitering phase.}
\label{f:desaceleracionB}
\end{figure}

In Figure \ref{f:desaceleracionB} we observe the behaviour of  Eq. (\ref{eq:q}), where we found that in the transition between the GR to Eddington's regime, the cosmic expansion is accelerating in the loitering phase around a minimum size before the GR-like expansion. Interesting enough, in this scenario we are obtaining an accelerating effect without the inclusion of a cosmological constant.

\begin{figure}
 \centering
  \subfloat[$\rho=1.1$]{
   \label{f:fig1.1}
    \includegraphics[width=0.4\textwidth]{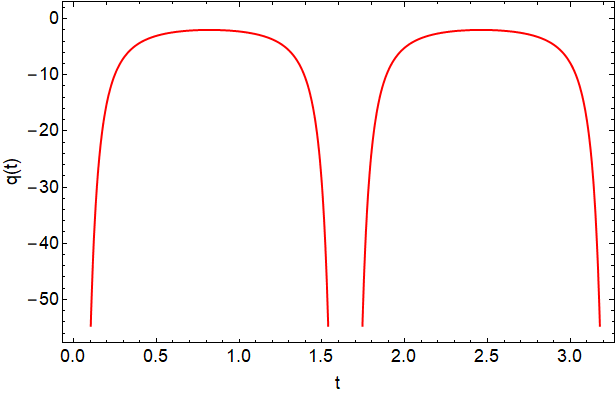}} \\
  \subfloat[$\rho=11$]{
   \label{f:fig1.2}
    \includegraphics[width=0.4\textwidth]{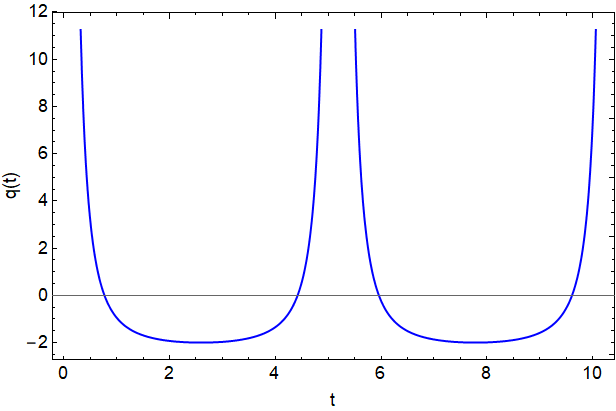}}
 \caption{Deceleration parameter for $\rho=1.1$ (top) and $\rho=11$ (bottom). These values for the density were chosen in such a way that we could observe the dynamics of the cosmic expansion in terms to the change in the value of the density.}
 \label{f:desaceleration}
\end{figure}

Now, if we compute the spatial components for the field equations we can obtain
\begin{equation}\label{eq:2}
\frac{\ddot{a}}{a}=-\frac{1}{\kappa}\left(1+\frac{D}{\kappa p-\lambda}\right)-2\left(\frac{\dot{a}}{a}\right)^2,
\end{equation}
where $D=XY^3$. Replacing Eq.(\ref{eq:2}) in Eq.(\ref{eq:1}) we obtain
\begin{eqnarray}
q & = &
% \left[\frac{1}{\kappa}\left(1+\frac{D}{\kappa p-\lambda}\right)+2\left(\frac{\dot{a}}{a}\right)^2\right]\left(\frac{\dot{a}}{a}\right)^{-2} \nonumber \\
\frac{1}{\kappa}\left(\frac{a}{\dot{a}}\right)^{2}\left(1+\frac{D}{\kappa p-\lambda}\right)-2.
\end{eqnarray}
For the GR regime we have $\lambda=1=D$, therefore
\begin{equation}
q=\frac{p}{p-1}\left(\frac{a}{\dot{a}}\right)^{2}-2.
\end{equation}
In the case of radiation $p=\rho /3$ with $\kappa=1$ we get
\begin{equation}
q(\rho,t)=\frac{\omega^{2} \rho}{\rho-3}\cot^{2}{\omega t}-2,
\end{equation}
with $\omega=\sqrt{\rho/3(\rho-1)}$. 
In Figure \ref{f:desaceleration} we notice two different behaviours: for $\rho=1.1$ we have a scenario where the universe expands as time evolves and with a deceleration parameter that decreases until it reaches a fixed point where it expands rapidly. For $\rho=11$ notice that the universe first decelerates, then accelerates which after a certain time it decelerates again: a bouncing universe with a minimum size. 

\begin{figure}[ht!]
 \centering
    \includegraphics[width=0.5 \textwidth]{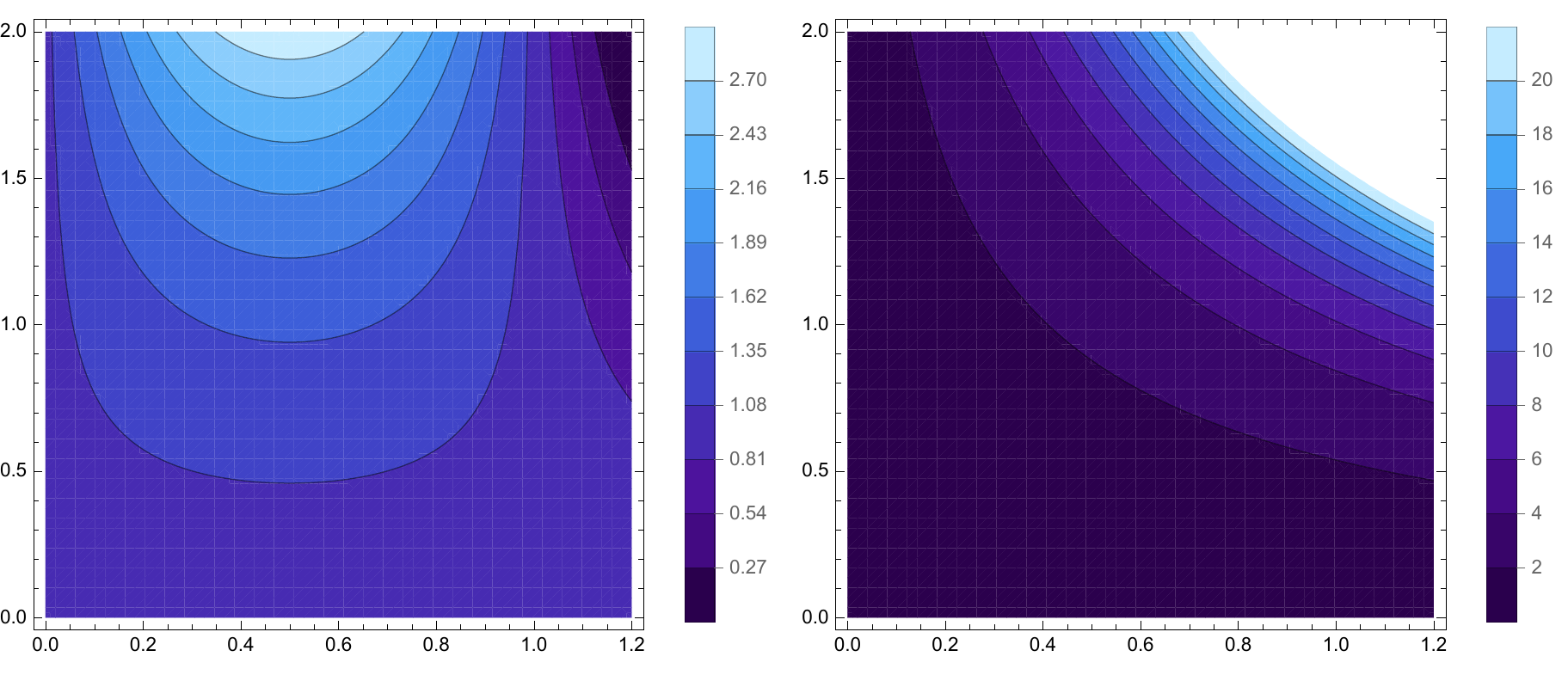}
 \caption{Contour plot for the parameter space $[\rho,t]$. \textit{Left:} The scale factor present a higher response value in the blue-smooth region, which represent the maximum value for $\kappa >0$. \textit{Right:} The scale factor present at least two higher response values in the blue-smooth regions, which represent a bouncing solution for $\kappa < 0$. }
 \label{f:scalefactor}
\end{figure}

\begin{figure}[ht!]
 \centering
    \includegraphics[width=0.5 \textwidth]{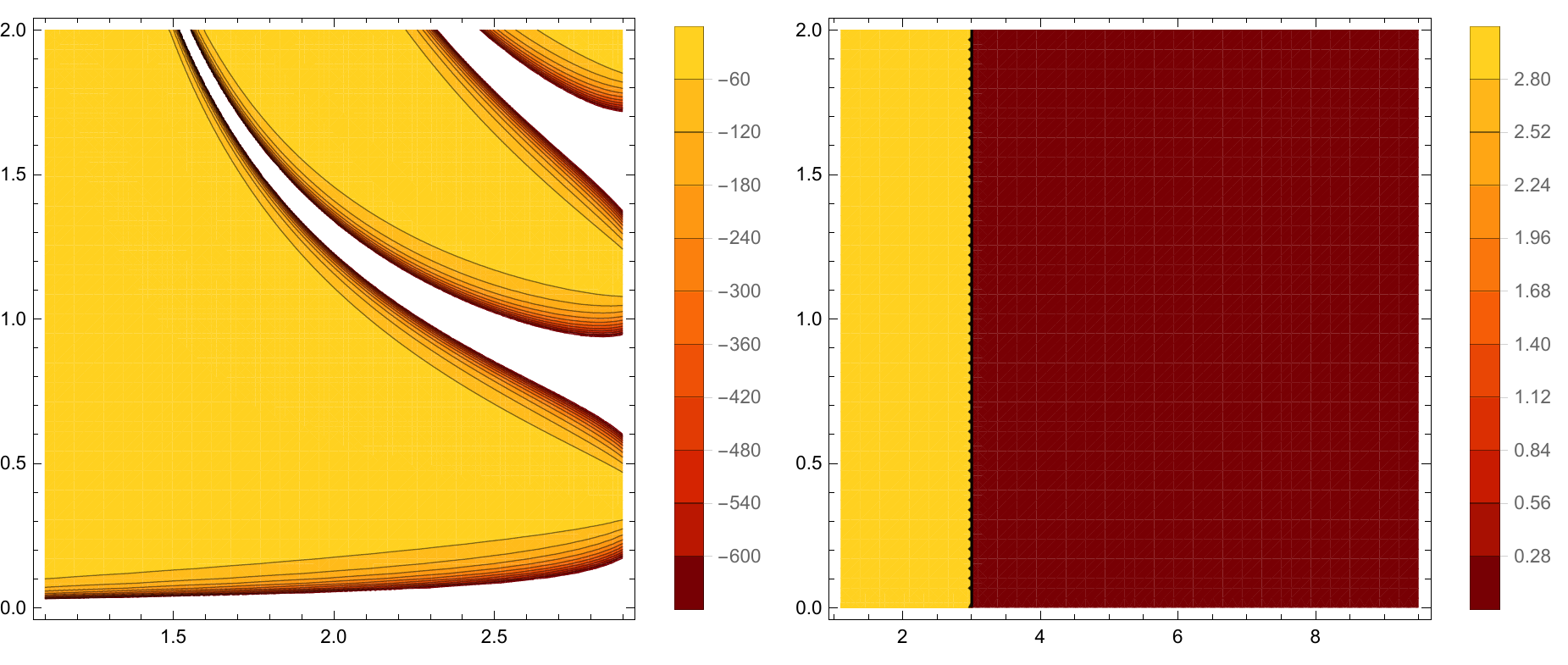}
 \caption{Contour plot for the parameter space $[\rho,t]$. \textit{Left:} The deceleration parameter present a higher response value in the yellow-smooth region, which represent a EiBI universe always accelerating for $\kappa >0$. \textit{Right:} The deceleration parameter present at least two response values in the darker blue regions, which represent a braking point in each bouncing solution for $\kappa < 0$. }
 \label{f:deceleration}
\end{figure}

\begin{figure}[ht!]
 \centering
    \includegraphics[width=0.5 \textwidth]{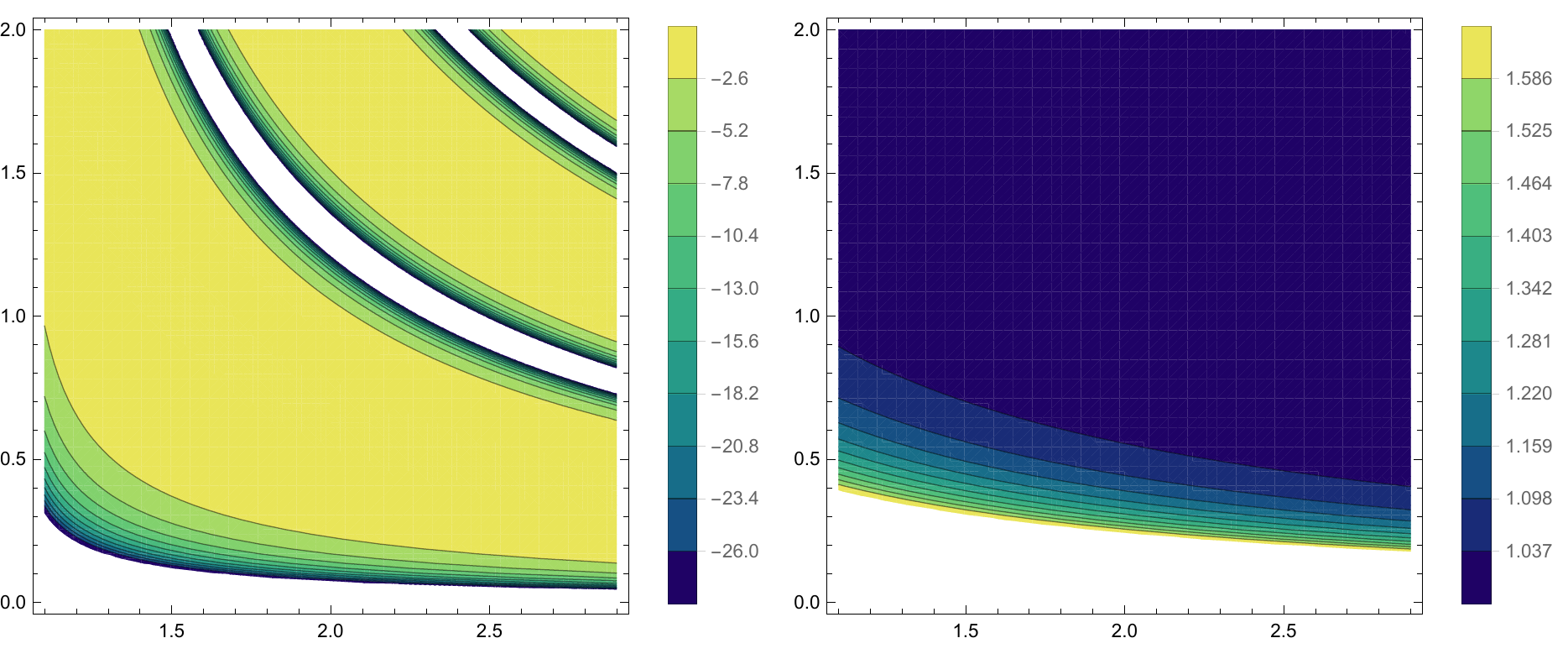}
 \caption{Contour plot for the parameter space $[\rho,t]$. \textit{Left:} Negative jerks implies linear decrease of acceleration for $\kappa >0$ for each smooth color region, which represent a deceleration build-up phase. \textit{Right:} Positive jerk values for $\kappa <0$, which represent a deceleration ramp-down in the darker region.}
 \label{f:jerk}
\end{figure}

\begin{figure}[ht!]
 \centering
    \includegraphics[width=0.5 \textwidth]{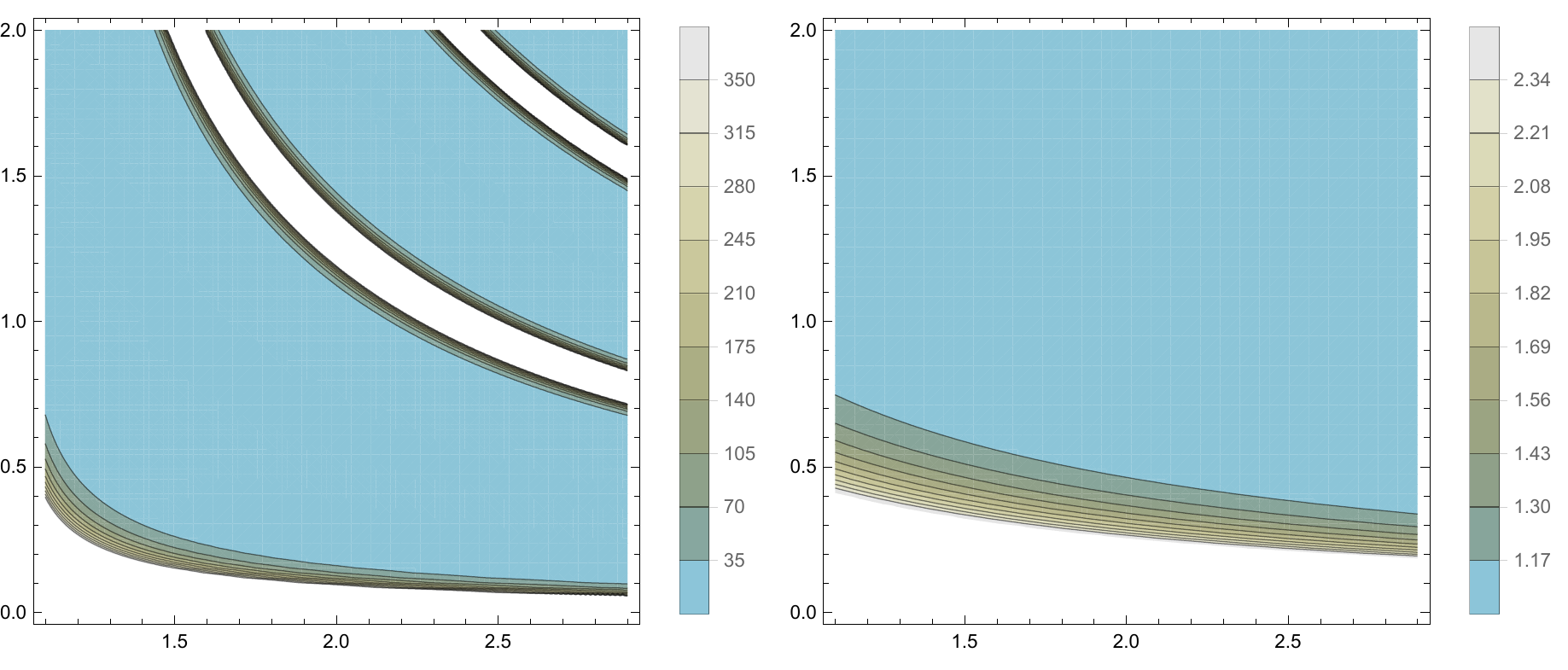}
 \caption{Contour plot for the parameter space $[\rho,t]$ for the snap and lerk parameters \textit{Left:} $\kappa >0$. \textit{Right:} $\kappa < 0$. Both parameters assume only positive values without any redshift transition.}
 \label{f:LS}
\end{figure}

With the latter analysis and using our Friedmann solution (\ref{eq7}), a cosmographic approach can be develop in order to constrain the parameters. The advantage of cosmography is the particularity to avoid any kind os form over the Friedmann equations which makes this analysis a model independent. Some calculations in regards to EiBI scenario has been presented in \cite{Bouhmadi-Lopez:2014jfa}, showing the presence of scenarios where the singularity still remains and the neccessity of high order cosmographic parameters are required. In this section we demonstrate that the singularities are a property that still can be avoided for the two values of $\kappa$ and a dark energy behaviour can be observed. To start we will use the standard cosmographic definitions:
\begin{eqnarray}
j(t) &=& \frac{1}{a} \frac{d^3 a}{dt^3}\frac{1}{H^3},
%-\cot^{2}{\omega t}, 
\\
s(t)&=&  \frac{1}{a} \frac{d^4 a}{dt^4}\frac{1}{H^4},
 %= \cot^{4}{\omega t}, 
 \\
l(t)&=&  \frac{1}{a} \frac{d^5 a}{dt^5}\frac{1}{H^5}, 
%= \cot^{4}{\omega t},
\end{eqnarray}
where $j$, $s$ and $l$ are the so-called jerk, snap and lerk parameters. For our EiBI scenario we obtain
\begin{eqnarray}
q+2 &=&-\frac{\omega^2 \rho}{\rho -3} j, \\
 s &=& l = j^2,
\end{eqnarray}
The evolution of each parameter, including the scale factor are showed in Figures \ref{f:scalefactor}-\ref{f:deceleration}-\ref{f:jerk}-\ref{f:LS}. Notice how the scale factor for the scenario with $\kappa > 0$ shows one high value, meanwhile for $\kappa < 0$ a series of bouncing solutions are detailed. The contour regions for each cosmographic parameters show a running along isoclines which corresponds to staying at a constant (flat) value, and crossing the isoclines corresponds to either ascent or descent of each of their values, respectively.

%%%%%%%%%%%%%%%%%%%%%%%%%%%%%%%%%%%%%%%%%%%%%%%%%%%%
%%%%%%%%%%%%%%%%%%%%%%%%%%%%%%%%%%%%%%%%%%%%%%%%%%%%

\newpage
\section{Conclusions}
\label{sec:conclusions}
Eddington-inspired-Born-Infeld theory (EiBI) continues being studied since its characteristic by being equivalent to Einstein theory in vacuum but differing from it in the presence of matter leads to features with the ability to avoid some singularities such as the Big Bang singularity in the finite past of the universe, and the singularity formed after the collapse of a star and in the presence of a maximum density $\rho_{B}$ in cosmology can influence not only the early universe, also in the dynamic formation of black holes \cite{Pani:2012qd,Avelino:2012ge,Tavakoli:2015llr}. Therefore, if a minimum length also arises during the gravitational collapse, the universe can be completely free of singularity, solving one of the problems that have troubled relativists since Einstein first proposed his theory of gravity. We found that also the deceleration parameter for this theory shows a accelerating cosmic expansion for both cases $\kappa=1$ and $\kappa=-1$, this in some sense can mimic the behaviour of a dark energy fluid in a scenario where a loitering phase is being carried out. 

%%%%%%%%%%%%%%%%%%%%%%%%%%%%%%%%%%%%%%%%%%%%%%%%%%%%
%%%%%%%%%%%%%%%%%%%%%%%%%%%%%%%%%%%%%%%%%%%%%%%%%%%%

\begin{acknowledgments}
N.M. Jim\'enez is supported by a CONACyT scholarship. CE-R is supported by the Royal Astronomical Society as FRAS 10147. 
\end{acknowledgments}

%%%%%%%%%%%%%%%%%%%%%%%%%%%%%%%%%%%%%%%%%%%%%%%%%%%%
%%%%%%%%%%%%%%%%%%%%%%%%%%%%%%%%%%%%%%%%%%%%%%%%%%%%

\end{document}